\documentclass{PoS}
\usepackage[]{units}

\title{Low Energy Cosmic Ray Spectrum from 250 TeV to 10 PeV using IceTop}

\ShortTitle{Cosmic Ray Energy Spectrum}

\author{
The IceCube Collaboration\footnote{For collaboration list, see PoS(ICRC2019) 1177.}\\
{\itshape \href{http://icecube.wisc.edu/collaboration/authors/icrc19_icecube}{http://icecube.wisc.edu/collaboration/authors/icrc19\_icecube}}\\
E-mail: \email{rkoirala@udel.edu, gaisser@udel.edu}
}

\abstract{
Using IceTop, the surface component of the IceCube Neutrino Observatory, the all-particle cosmic ray energy spectrum has been determined above a few PeV. The measured energy spectrum leaves a gap of more than a decade in energy to direct measurements by balloon and satellite experiments. In this analysis, we lowered the energy threshold of IceTop to 250 TeV, narrowing the gap between IceTop and direct measurements. In order to collect lower energy events, we implemented a new trigger that uses four pairs of infill stations for which the separation between stations is less than 50 m, compared to 125 m for the main array. The new trigger collects data from the entire array for events with hits on at least one infill pair. The low-energy extension of the all-particle cosmic ray energy spectrum using these IceTop events is measured and is compared with the energy spectrum from HAWC and other experiments. Air shower simulations with two different hadronic interaction models, Sibyll 2.1 and QGSJetII-04, are used in this analysis and an energy spectrum for each model is produced. Both measured energy spectra show a bend around the knee region.\\

\vspace{4mm}
{\bfseries Corresponding authors:}
\speaker{Ramesh Koirala}$^{1}$, Thomas K. Gaisser$^{1}$\\
{$^{1}$ \itshape Bartol Research Institute, Dept.~of Phys.~and Astr., Univ.~of Delaware, Newark, DE, U.S.A.}\\

}

\FullConference{36th International Cosmic Ray Conference -ICRC2019-\\
		July 24th - August 1st, 2019\\
		Madison, WI, U.S.A.}

\begin{document}

\section{Introduction}\label{sec:intro}

    Cosmic rays are energetic charged particles that reach Earth from space. The cosmic ray energy spectrum is an important tool to study acceleration and propagation of cosmic rays. Using satellite and balloon experiments, the cosmic ray spectrum and its chemical composition are directly measured up to \unit[100]{TeV}. Above \unit[100]{TeV}, they are indirectly measured by air shower experiments. The IceTop spectrum thus far covers an energy region above \unit[1.58]{PeV}~\cite{bhaktiyar,IceCube_3year_composition_2019}. In this study, we measure the cosmic ray energy spectrum from \unit[250]{TeV} to \unit[10]{PeV}. This reduction of the energy threshold has almost connected the IceTop measurements with the direct measurements. 
    
\begin{figure}[t]
	\centering 
	\includegraphics[height=6cm]{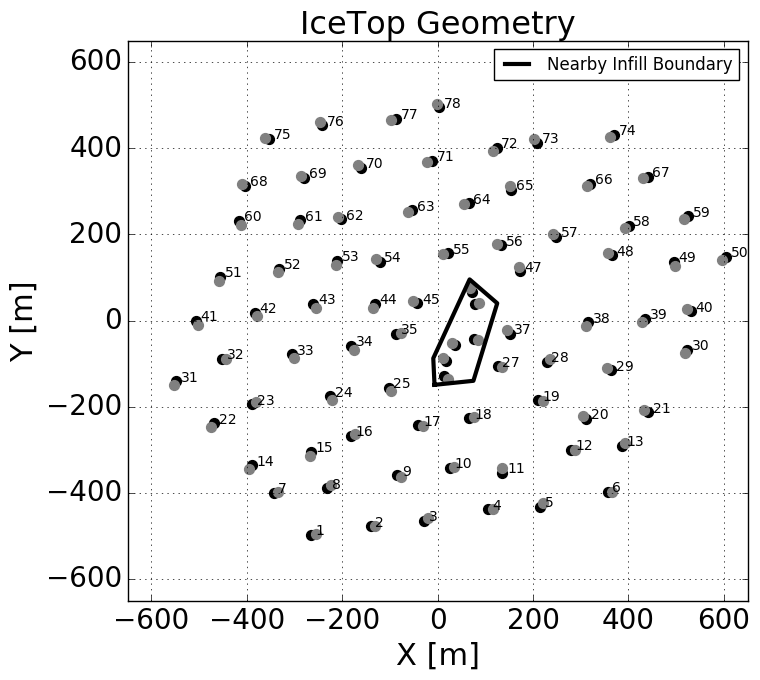}
	\includegraphics[height=6cm]{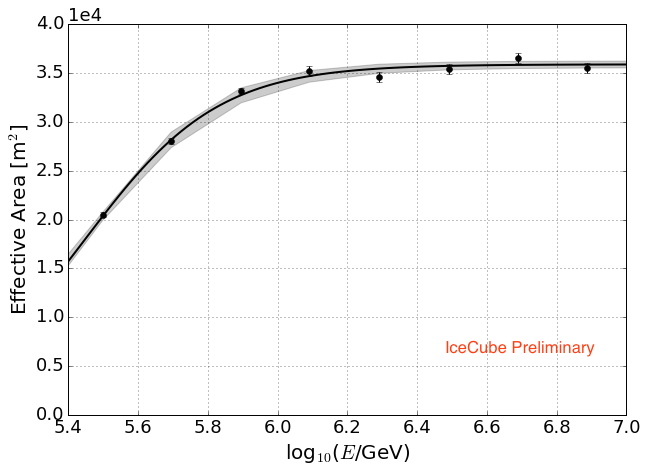}
	\caption{Left: IceTop geometry with positions of all tanks. Nearby infill boundary includes 6 out of 8 infill stations. Tanks inside this boundary at the center are used to define a two-station trigger. Right: Effective area calculated using MC generated with modified H3a composition model, Sybill 2.1 hadronic interaction model, and cosine of zenith angle greater than or equal to 0.9. The line is a fit through the simulated points.}
	\label{it_geo_aeff}
    \end{figure}

    IceTop is the surface component of the IceCube Neutrino Observatory~\cite{icetop_detector}. It consists of 162 tanks distributed in 81 stations spread over an area of \unit[1]{km$^2$} as shown in the left plot of Fig \ref{it_geo_aeff}. Each station has two tanks separated by \unit[10]{m}. Stations are arranged in a triangular grid and two nearby stations are approximately \unit[125]{m} apart. In addition, IceTop has a dense infill array where the distance between nearby stations is smaller than \unit[125]{m}. The dense infill array is used to detect cosmic rays with comparatively lower energy. 
    
    Data collected by IceTop are primarily used to study the cosmic ray energy spectrum~\cite{bhaktiyar,IceCube_3year_composition_2019,it26_spectrum,ic40_spectrum,Rawlins:2016bkc} and the mass composition of primary particles~\cite{IceCube_3year_composition_2019}. To lower the energy threshold of IceTop to \unit[250]{TeV}, a new trigger and filter was developed and has been implemented since May 20, 2016. The trigger identifies small showers which hit at least two stations in the infill array. This paper reports the all-particle cosmic ray energy spectrum using these two-station events collected from May 2016 to April 2017 with a livetime of 330.43 days. There are 7,442,086 number of two-station events after all quality cuts.

\section{Trigger and Filter}\label{section_detector}
    A station is triggered if both tanks are hit within a certain time interval (local coincidence). A two-station trigger is formed by counting the number of tanks from 6 infill stations within a cylinder defined by a \unit[60]{m} radius that are hit within a \unit[200]{ns} time window. 4 pairs of stations can be formed from 6 individual infill stations with a  distance of less than \unit[60]{m} between each station. The trigger condition is satisfied if any pair of stations is hit. Once the trigger condition is fulfilled, all other local coincidences within 10$\mu$s before and after the first and last hit are included. All triggered events automatically pass the filter condition and are sent to the North.
    
\section{Analysis}\label{section_analysis}
    A random forest regression~\cite{Breiman:2001hzm, James:2014} is used separately to reconstruct the air shower's core position, zenith angle, and energy. A Monte Carlo simulation is used to train the random forest and develop a prediction model. Reconstruction of the core position includes reconstructing x and y coordinates of the core separately. The major features that are used to reconstruct the shower's core are position of the charge center of gravity, position of hit tanks, and the charge on hit tanks. The same technique is used for training and predicting zenith angle. The major features that are used to reconstruct the zenith angle are direction assuming a plane shower front, the time of the hits, and the average distance of tanks that have been hit from the plane shower front. The energy reconstruction uses reconstructed shower core and zenith angle. The major features for an air shower's energy reconstruction are the charge deposited on tanks, distance of hit tanks from the shower core, and the zenith angle of an air shower. Charge is calibrated in units of vertical equivalent muons (VEM)~\cite{icetop_detector}.

	\begin{figure}[t]
	\centering
	\includegraphics[height=4.9cm]{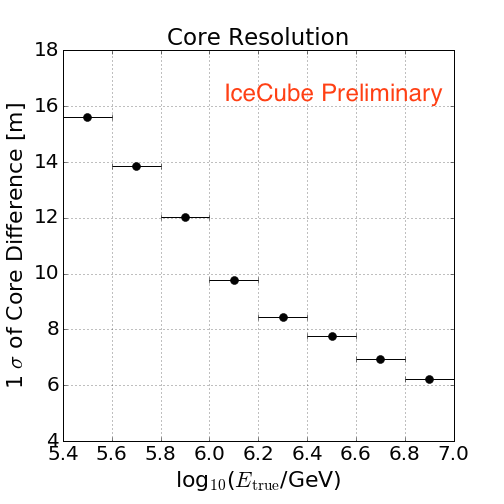}
	\includegraphics[height=4.9cm]{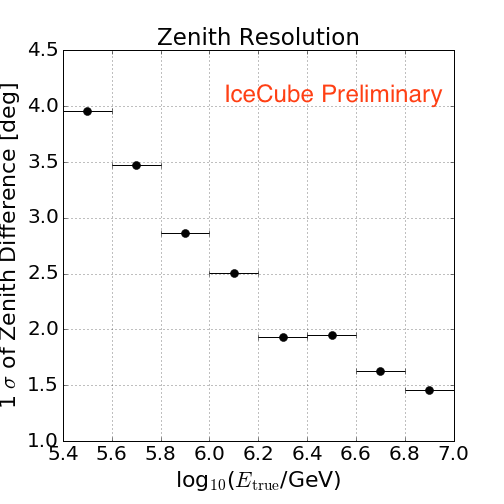}
	\includegraphics[height=4.9cm]{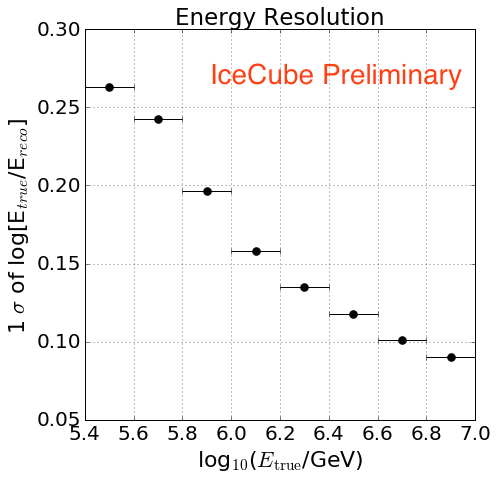}
	\caption{Left: Core resolution in meter; Middle: zenith resolution in degree; Right: energy resolution in unit-less quantity.}
	\label{resolution_plot}
	\end{figure}

    Fig \ref{resolution_plot} shows core resolution, zenith resolution, and energy resolution. The core resolution is about \unit[16]{m}, the zenith resolution is about $4^\circ$, and the energy resolution is about 0.26 for the lowest energy bin ($\log_{10}(E/{\rm GeV})$ 5.4 to 5.6). All three resolutions improve as energy increases.
    
    Only well-reconstructed events are used to obtain the energy spectrum. Quality cuts are used to remove events with possible bad reconstruction to reduce error and to improve accuracy. The passing rate of events for a cut is the percentage of events surviving that cut and all previous cuts. The following cuts are applied to the simulated and the experimental data to select events:.
    \begin{itemize}
	\item Events must pass the two-station trigger and filter. Passing rate for this cut is 100\%.
	\item Events must have the tank with the highest charge inside the nearby infill boundary. This cut is designed to select events with shower core near the nearby infill boundary. Passing rate for this cut is 89.5\%.
	\item Events must have cosine of zenith angle ($\theta$) greater than or equal to 0.9. These events have higher triggering efficiency and are better reconstructed. Passing rate for this cut is 48.1\%.
	\item Events with most of the energy deposited only in few tanks are removed, as they are known to be poorly reconstructed. This cut requires the largest charge to be less than or equal to 75\% of the total charge and the sum of the two largest charges less than or equal to 90\% of the total charge. Passing rate for this cut is 36.836\%.
	\item The simulation used for this low energy analysis extends to log$_{10}$($E$/GeV)=7.4. From the simulation we have determined that events with a true energies above that can be removed by excluding events with more than 42 stations hit and excluding events with a total charge greater than \unit[10$^{3.8}$]{VEM}. Passing rate for this cut is 36.835\%.
    \end{itemize}

    An iterative Bayesian unfolding~\cite{DAgostini1995, dagostini_unfolding} is used to take energy bin migration into account and to derive the true energy distribution (d$N$/dlog$_{10}(E$/GeV)) from the reconstructed energy distribution. It is implemented via a software package called pyUnfolding~\cite{pyunfolding}. This package also calculates and propagates error in each iteration.
    
    To unfold the energy spectrum, the response of the detector to an air shower is required. The response is determined from simulations. This information is stored in a response matrix and is the probability of measuring a reconstructed energy given the true primary energy. Inverting the response matrix to get a probability of measuring true energy given reconstructed energy would lead to unnatural fluctuations. Therefore, the Bayes theorem is used iteratively to get a true distribution from an observed distribution. The unfolding proceeds until a desired stopping criterion is satisfied. In this analysis, a Kolmogorov-Smirnov test~\cite{kolmogorov,smrinov} of the subsequent unfolded energy distribution less than $10^{-3}$ is used as the stopping criterion. The choice of an initial prior, $P(E)$, is optional and can be any reasonable distribution. An initial prior used in this analysis is Jeffreys' Prior~\cite{jeffreys_prior}. The unfolded energy spectrum after each iteration is used to update the prior for the next iteration. The prior can quickly become unphysical if it is not regularized after each iteration. So, a spline fit is performed on the unfolded energy spectrum to regularize the updated prior.

    The rate of events fluctuates with changes in atmospheric pressure. If pressure increases, the rate decreases and vice-versa. If the average pressure during which data were taken is not equal to the pressure of atmospheric profile used in the simulation, then the final flux must be corrected to account for the difference in the atmospheric pressure between data and simulation. The average pressure at the South Pole during data-taking was \unit[691.16]{g/cm$^2$} (data obtained from the Antarctic Meteorological Research Center). For the energy region of interest in this analysis, the atmosphere used in simulation has an average pressure of \unit[698.12]{g/cm$^2$}. The effect of this pressure difference on the final flux is corrected by using data from a time period when the atmospheric pressure is approximately \unit[698.12]{g/cm$^2$}. The correction factor is calculated by comparing this flux with the flux using all data and is around 5\%. 

\subsection{Systematic Uncertainties}
\begin{figure}[t]
	\centering
	\includegraphics[height=5.8cm]{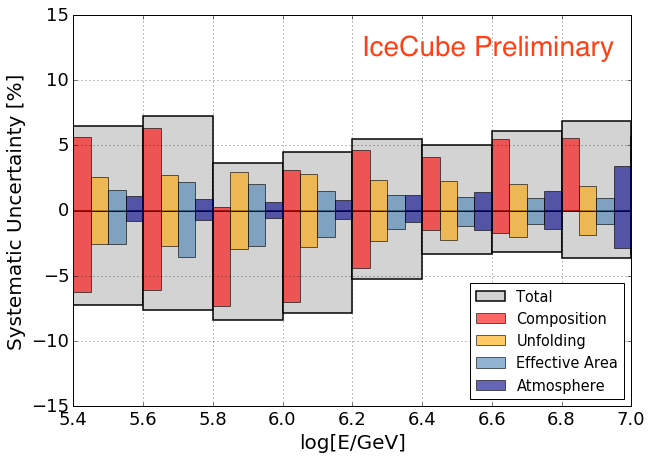}
	\caption{The individual systematic uncertainties for each energy bins. Total systematic uncertainty is the sum of individual uncertainties added in quadrature.}
	\label{total_sys_uncertainty}
    \end{figure}

    Systematic uncertainties are calculated by keeping all conditions constant except the feature under investigation. The systematics uncertainties due to the hadronic interaction models is considered separately. The major systematic uncertainties, excluding those due to the hadronic interaction models, are those due to the composition, the unfolding method, the effective area, and the atmosphere. Individual and `total systematic uncertainty' are shown in Fig \ref{total_sys_uncertainty}.
    
    To know the uncertainty due to composition, the modified Gaisser H3a model is used as a base composition model and GST~\cite{Gaisser:2013bla}, GSF~\cite{Dembinski:2017zsh}, and a modified version of Polygonato are used as alternate composition models. Since all these models are viable options for composition, the flux for each model is calculated using the same response matrix and the percentage deviation of the flux from the model for each energy bin is measured. Additionally, the fractional difference between fluxes calculated for two extreme zenith bins is used to calculate composition systematics as done for the 3-year energy spectrum analysis. The maximum spread of all deviations is used as the uncertainty due to composition.

    The pyUnfolding package calculates the systematic uncertainty due to unfolding after each iteration. The uncertainty arises from limited amount of statistics of the simulated data set. In this study, we need 12 iterations before reaching the termination criterion. The systematic uncertainty for the twelfth iteration is used as the systematic uncertainty due to the unfolding procedure.
    
    The effective area is fitted with an energy-dependent sigmoid function. The parameters of the fit contain errors and the errors have to be accounted for while calculating the flux. A band around the effective area fit is shown in the right plot of Fig. \ref{it_geo_aeff} after accounting for all errors on the parameters. The effective area and the error band shown in the figure is for events with cosine of zenith angle greater than or equal to 0.9. Taking the upper and lower boundary of the band, the flux is calculated and the difference in the flux is used as the systematic uncertainty due to the effective area.
    
    The uncertainty on the correction factor to account for the atmospheric pressure difference between experimental data and simulation is used as the systematic uncertainty due to pressure. Also, the difference in flux due to different temperatures for constant pressure is used as the systematic uncertainty due to temperature and is less than 2\%. These two uncertainties are added and the summation is used as the systematic uncertainty due to the atmosphere.
    
    Different snow heights for data and simulation affect the low energy spectrum analysis. Experimental data used in this analysis is from May 2016 to April 2017 and the snow height used for simulations is from October 2016. Since the average height of snow between experimental data and simulation is comparable, the systematic uncertainty due to snow is estimated to be small.
    
    The statistical uncertainty is small due to the large volume of data. The systematic uncertainty from the composition assumption is the largest, whereas, the systematic uncertainties from the unfolding method, effective area, and atmosphere are relatively small. The `total systematic uncertainty' is calculated by adding individual contributions in quadrature and is larger than the statistical uncertainty.

\section{Flux}
    Once the core position, direction, and energy of air showers are reconstructed, and the effective area is known, the cosmic ray flux, or energy spectrum, is calculated. The binned cosmic ray flux is given by
    \begin{equation}
    J(E) = \frac{\Delta N(E)}{\Delta {\rm ln}E \pi(\cos^2\theta_1 - \cos^2\theta_2) A_{\rm eff} T},
    \label{flux_equaiton_final}
    \end{equation}
    where $\Delta N(E)$ is the unfolded and pressure corrected number of events with energy per logarithmic bin of energy in time $T$, [$\theta_1, \theta_2$] is the observed zenith range, and $A_{\rm eff}$ is the effective area. The effective area for IceTop events with $\cos\theta\geq 0.9$ is shown in the right plot of Fig \ref{it_geo_aeff} and is used to calculate the flux. The livetime ($T$) used is \unit[28548809.85]{s} (330.43 days), $\Delta \log_{10}E$ used is 0.2, and $\cos\theta_1$ and $\cos\theta_2$ used are 1.0 and 0.9 respectively.
    
    \begin{figure}[t]
	\centering
	\includegraphics[height=5.8cm]{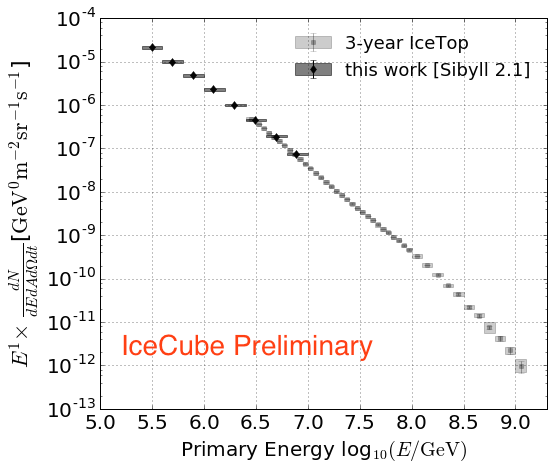}	\includegraphics[height=5.8cm]{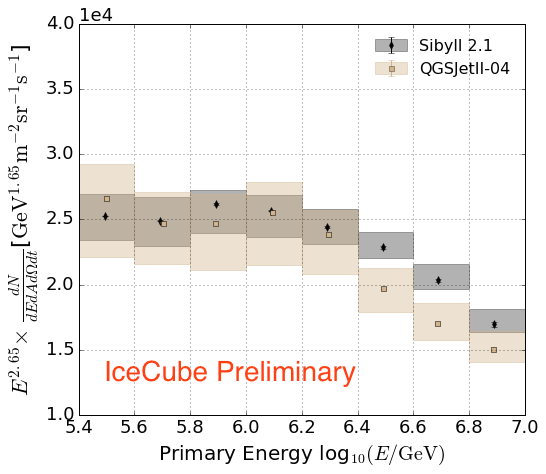}
	\caption{Left: The all-particle cosmic ray energy spectrum using IceTop 2016 data. The analysis is done using simulations with Sibyll 2.1 as the hadronic interaction model. Right: The all-particle cosmic ray energy spectra using simulations with Sibyll2.1 and QGSJetII-04 as hadronic interaction models. The same analysis as with Sibyll 2.1 was repeated with QGSJetII-04. The right plot is scaled by $E^{1.65}$. The shaded region in both plots indicates the systematic uncertainties.}
	\label{cr_flux}
    \end{figure}
    
    The all-particle cosmic ray flux is calculated using Eq. \ref{flux_equaiton_final} in the energy range \unit[250]{TeV} to \unit[10]{PeV} and compared with the higher energy measurement of IceCube~\cite{IceCube_3year_composition_2019} in the left panel of Fig \ref{cr_flux}. The spectrum is plotted per logarithmic bin of energy in units of $\rm m^{-2}s^{-1}sr^{-1}$.

    The effect of the hadronic interaction model is not included in the `total systematic uncertainty'. Instead, the same analysis steps were repeated using simulation with QGSJetII-04 as the hadronic interaction model. The statistics of the simulation for the analysis with QGSJetII-04 is only 10\% of that for Sibyll 2.1 but is sufficient for the comparison between the two models, as shown in the right plot of Fig \ref{cr_flux}.

    \begin{figure}[t]
	\centering
	\includegraphics[height=6.9cm]{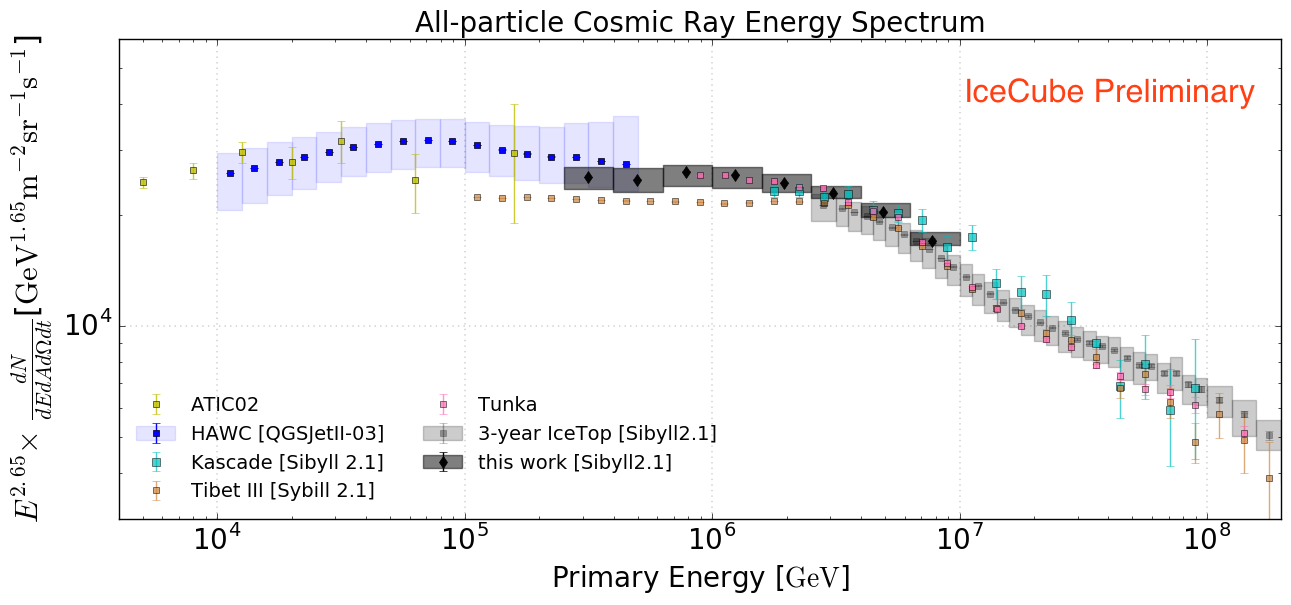}
	\caption{Cosmic ray flux using IceTop 2016 data scaled by $E^{1.65}$ and compared with flux from other experiments. This analysis and HAWC's energy spectrum analysis use different hadronic interaction models. The shaded region indicates the systematic uncertainties.}
	\label{cr_flux_many}
    \end{figure}

    Many ground-based cosmic ray detectors measured the cosmic ray flux around this energy region. Several measurements with their statistical uncertainties are compared with the result of this analysis in Fig \ref{cr_flux_many}. The range of fluxes reflects systematic uncertainties in the measurements. Since the cosmic ray flux follows a steep power law, a slight difference in energy scale can cause a large difference in the flux. The IceTop low energy spectrum extension overlaps with the results from HAWC~\cite{HAWC2017} in the lower energy region and with Kascade~\cite{kascade_spectrum} and Tunka~\cite{PROSIN201494} measurements at higher energy region, and is higher than the result from Tibet III~\cite{Amenomori}. The low energy spectrum is also compared with a direct measurement from ATIC-02~\cite{atic_spectrum}. Since the hadronic interaction model is different for HAWC and this analysis, this difference can affect their results.

\section{Discussion}\label{section_result}
    With this analysis the all-particle cosmic ray energy spectrum from \unit[250]{TeV} to \unit[10]{PeV} was measured, lowering the energy threshold of IceTop from $\sim$\unit[2]{PeV} to \unit[250]{TeV}. Most effort went into the deployment of a new trigger and filter to select low energy events and into the development of a new reconstruction method for these hard to reconstruct events. 

	The final energy spectrum from this analysis is shown in figures \ref{cr_flux} and \ref{cr_flux_many}. The right plot in Fig \ref{cr_flux} shows that the flux is higher than the 3-year IceTop spectrum in the overlap region. Both IceTop spectra were fitted using spline fits, and their percentage differences is within 8.5\%. The `total systematic uncertainty' obtained by adding the individual uncertainties in quadrature for the 3-year spectrum is 9.6\% at \unit[3]{PeV} and 10.8\% at \unit[30]{PeV}~\cite{bhaktiyar}. Even though the flux obtained with this analysis is higher than that of the 3-year analysis in the overlap region, both are compatible within their systematic uncertainties. Both analyses use data collected by IceTop, so they share systematic uncertainties related to the detector. However, in this analysis some systematics are differently treated, like the pressure correction and the unfolding. Other important differences are in data taking (trigger/filter) and in the use of machine learning for reconstruction.

	The distinct feature of the spectrum is a bend in the knee region as observed in the 3-year IceTop spectrum~\cite{IceCube_3year_composition_2019} and in spectra of many other experiments. Additionally, the energy spectrum overlaps with HAWC's result~\cite{HAWC2017} within the systematic errors around \unit[300]{TeV}. The energy spectrum measured in this analysis fills the gap between the 3-year IceTop spectrum and the HAWC measurements and thus connects with direct measurements.

\bibliographystyle{ICRC}
\bibliography{references}

\end{document}